# CLARA: A Developer's Companion for Code Comprehension and Analysis

Ahmed Adnan
*University of Dhaka*
Dhaka, Bangladesh
bsse1131@iit.du.ac.bd

Mushfiqur Rahman*
*Bangladesh University of Business & Technology*
Dhaka, Bangladesh
Mushfiqur.Rahman@bubt.edu.bd

Saad Sakib Noor*
*University of Dhaka*
Dhaka, Bangladesh
bsse1122@iit.du.ac.bd

Kazi Sakib
*University of Dhaka*
Dhaka, Bangladesh
sakib@iit.du.ac.bd

*Abstract*—Code comprehension and analysis of open-source project codebases is a task frequently performed by developers and researchers. However, existing tools that practitioners use for assistance with such tasks often require prior project setup, lack context-awareness, and involve significant manual effort. To address this, we present CLARA, a browser extension that utilizes state-of-the-art inference model to assist developers and researchers in: (i) comprehending code files and code fragments, (ii) code refactoring, and (iii) code quality attribute detection. We qualitatively evaluated CLARA's inference model using existing datasets and methodology, and performed a comprehensive user study with 10 developers and academic researchers to assess its usability and usefulness. The results show that CLARA is useful, accurate, and practical in code comprehension and analysis tasks. CLARA is an open-source tool available at *github.com/clara_tool_demo*. A video showing the full capabilities of CLARA can be found at *youtube.com/clara_demo_video*.

## I. INTRODUCTION

Open-source GitHub repositories provide rich sources of production-ready code, reusable code components, and valuable insights into software development practices [1, 2]. Developers and researchers frequently explore these codebases to understand the implementation details and functional behavior of the project code [1, 3]. While important, the scale and complexity of these repositories often pose a challenge in effective and efficient code comprehension and analysis.

To support developers and researchers in code comprehension and analysis tasks, various automated approaches have been proposed [4–7]. Researchers have also proposed several IDE-integrated and web-based tools [8–14] to support these tasks. However, most of these tools are designed to serve a single purpose, lack context-aware responses, and require prior project configuration or full project download. Practitioners also use generative AI models (e.g., chatGPT, Gemini), where they manually copy code and craft prompts to perform code comprehension and analysis, which requires considerable manual effort and often lacks repository context.

To address these limitations, we introduce CLARA, a browser extension that supports users in code comprehension and analysis. CLARA scrapes and parses code and contextual data from codebases and utilizes a state-of-the-art inference model to do (i) code explanation, (ii) code refactoring, and (iii) code quality attribute detection. CLARA also provides an AI-based chatbot to assist users performing these tasks. CLARA has modularized architectural components, well-designed APIs and can support multiple concurrent users, making it an extensible and scalable solution.

We evaluated CLARA in a user study with 10 participants, where more than 80% users found the tool easy to use, accurate, and helpful for code comprehension and analysis in GitHub codebases. We also conducted a comprehensive qualitative study to select the most suitable inference model for CLARA by evaluating multiple generative models' performance on three datasets relevant to CLARA's features. CLARA is available in Chrome Web Store [15] and can be installed easily on any Chromium-based browser (Chrome, Brave, Edge).

## II. CLARA: A CODE EXPLAINER & REPO ANALYZER

CLARA is a browser extension that assists developers and researchers in exploring and analyzing GitHub repository codebases. CLARA supports three code comprehension and analysis tasks, and features an AI-powered chatbot to assist users in performing these tasks.

### A. Supported Code Explanation & Analysis Tasks

*1) Code Explanation:* CLARA summarizes a code file with detailed descriptions of its attributes, methods, and context within the project repository. Additionally, the tool can explain any selected/highlighted portion of code and its purpose within the context. This feature provides an easier and efficient code comprehension experience.

*2) Code Refactoring:* When a user browses a code file in a GitHub repository, CLARA suggests a clean and refactored version of that source code, annotated with descriptive comments highlighting the applied changes. This feature makes the code components more reusable, frequently sought by open-source developers and researchers.

*3) Code Quality Attribute Detection:* CLARA detects code quality attributes such as cyclomatic complexity, maintainability index [16], and CVE vulnerability classification [17] of a browsed code file. This allows users to make a quick assessment of code quality without manually inspecting the code, identify high-impact areas and potential vulnerabilities.

In addition, CLARA provides an AI-based chatbot to assist users in follow-up inquiries. This chatbot supports all of CLARA's features, allowing users to seamlessly interact with the tool and receive accurate, context-aware responses.

---

*Both Mushfiqur Rahman and Saad Sakib Noor contributed equally as 2nd authors.

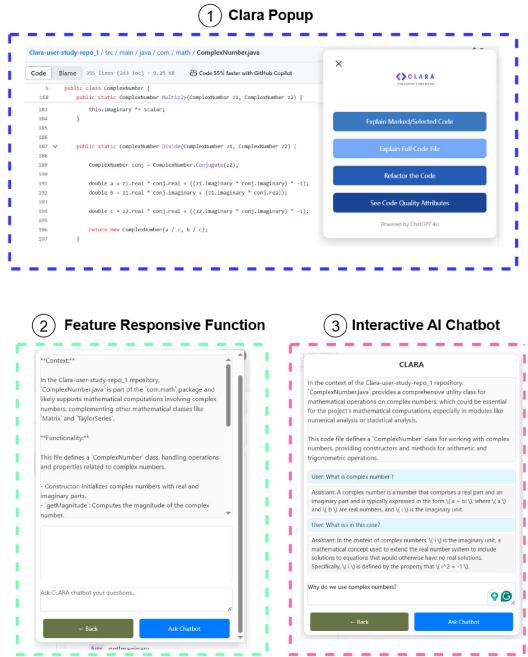

Fig. 1: CLARA's Usage Scenario

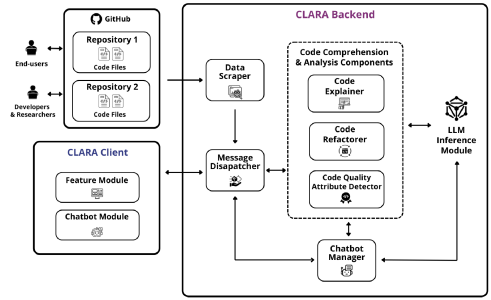

Fig. 2: Overview of CLARA's Architecture

### B. Usage Scenario & Graphical User Interface

CLARA can be easily installed in any Chromium-based browser(Chrome, Brave, Edge) via its Chrome Web Store page [15] by clicking 'Add to Browser'. Once installed, visiting any GitHub code file triggers CLARA's pop-up at the top-right (see ① in Fig. 1)

In CLARA's popup, there are four buttons. The "Explain Full Code File" button can give a context-aware explanation of the code file. The "Explain Marked/Selected Code" button can help the user to understand a specific highlighted part. The "Refactor the Code" button provides a refactored version of the source code with descriptive comments. The 'See Code Quality Attributes' button displays quality metrics of the code file. After clicking on any of these 4 buttons, CLARA's generated response (② in Fig. 1) is displayed in a module. A user can also leverage an AI-assisted chatbot (③ in Fig. 1) while using each of CLARA's features and ask followup questions or inquiries in a continuous conversation.

### III. CLARA 'S ARCHITECTURE & IMPLEMENTATION

CLARA's architecture, shown in Fig. 2, is composed of two subsystem units: (1) Backend, and (2) Client.

### A. Backend of CLARA

CLARA's backend consists of four main components:

*1) Data Scraper:* After CLARA is installed in a browser, when a user visits the code files of an open-source GitHub repository, this component scrapes necessary code file information along with repository metadata essential for contextual understanding (e.g., repo title, file tree containing paths of all files and folders, readme information, tags and topics) and communicates with message dispatcher for data propagation.

*2) Message Dispatcher:* This component acts as a middleware between CLARA's client and backend. This component dispatches client-side request to the relevant code comprehension/analysis component or to the chatbot manager. Likewise, the generated feedback response is sent back to the client's appropriate module from the backend through this component.

*3) Code Comprehension & Analysis Components:* There are three components in CLARA that assist users with code comprehension and analysis tasks :

Code Explainer: This component retrieves the necessary code file data, repository's contextual information and the required action event from the message dispatcher. Then, it constructs a well-defined prompt by parsing the retrieved code and repository context data and sends it to the LLM inference module. After receiving the response, the component processes the output, formats it appropriately, and returns it using message dispatcher.

Code Refactorer: When a user requests code refactoring, this component receives the action request and the code file data from message dispatcher, constructs an appropriate prompt with instructions to generate a clean, refactored version of the code with descriptive comments about refactoring changes, passes it to the LLM inference module. Then, it returns the formatted response to the client via message dispatcher.

Code Quality Attribute Detector: This component analyzes the code file data and predicts quality attributes, such as cyclomatic complexity, maintainability index, and CVE classified vulnerabilities [17]. Upon receiving the action request from the message dispatcher, this component breaks the code, formats it properly, and constructs a suitable prompt to for the inference module. Then, it utilizes the message dispatcher to return the formatted output.

Chatbot Manager: The chatbot manager facilitates communication between the client's chatbot module and backend's LLM inference module. In doing so, it manages conversational history, parses relevant information (both from scraped data and user inquiries), and constructs well-formatted and structured prompts and response messages accordingly.

### B. CLARA's Client Side (Frontend)

CLARA's client design structure contains a popup with four buttons (① in Fig. 1), representing its features. Additionally,

CLARA's feedback response section is divided into two key modules: (1) Feature Module, and (2) Chatbot Module. The feature module shows the generated feedback response in a structured and user-friendly format. The Chatbot Module contains an input field for receiving user inquiries and an output area where both user questions and chatbot responses are displayed in a clear, conversational format.

### C. CLARA's Extensibility & Scalability

CLARA's architectural components are well-modularized as its feature components are well-encapsulated from scraper and event handler components. CLARA also conforms with Chrome Extension Developer Guidelines [18] to enable easy extensibility. Though CLARA is built for comprehending and analyzing GitHub codebases, most of its backend components can be easily adapted for other platforms that contain project codebases.

CLARA's backend leverages Javascript's event-driven architecture and async/await-based non-blocking I/O system. This allows the backend to simultaneously handle multiple user requests without blocking the event loop. Additionally, since CLARA is deployed on the Chrome Web Store [15], each user's browser runs on an isolated instance that ensures client-side concurrency. This makes CLARA a scalable solution for users to perform code comprehension and analysis tasks.

### D. CLARA's Implementation Details

CLARA's client and backend are implemented using JavaScript. The client-side UI is designed using HTML and CSS. CLARA is developed as a browser extension, compatible with Chromium engine-based browsers (e.g., Chrome, Brave, Edge) and is deployed at Chrome Web Store [15]. CLARA utilizes Chrome and GitHub APIs to scrape data from GitHub repositories and uses the base GPT-4o [19] model through OpenAI's API as inference module.

## IV. CLARA'S EVALUATION

### A. Model Evaluation & Selection

To select an appropriate inference model for CLARA's features, we chose 3 datasets, resembling CLARA's 3 features for model evaluation. We selected the *CodeSearchNet* dataset [20] by GitHub for code comprehension, *MaRV* dataset [21] proposed by Nunes et al. for code refactoring, and *Big-Vul* dataset proposed by Fan et al. [22] for code quality attribute detection. After dataset selection, we applied random sampling to extract 30 samples from each dataset to evaluate the models' performance on them qualitatively. Then we carefully created prompts for each of the tasks using persona prompting and adhered to prompt engineering guidelines proposed by Ronaki et al. [23]. As for model selection, we chose 3 widely-used benchmark generative models (GPT-4o, Gemini 2.5 Flash, and DeepSeek v3. for their superior performance in LiveBench [24] benchmark test. After that, each author manually reviewed each of the models' responses, compared them with datasets' sample answers and qualitatively rated them. Finally, GPT-4o model [19] was selected due to it receiving the highest average rating. The evaluation details and data are given in our replication package [25]

### B. User Study

Since code comprehension and analysis tasks in GitHub codebases are frequently done by both industry practitioners and academic researchers, we conducted a user study involving 5 professional developers and 5 academic researchers. The participating academic researchers were from College of William & Mary and IIT, University of Dhaka and the developers were from some reputed software industries in Bangladesh, with an overall experience range of 2 to 8 years. The goal of the study was to evaluate (1) CLARA's usability/usefulness (**RQ1/RQ2**), (2) the quality and predictive accuracy of CLARA's suggestions (**RQ3**), and (3) CLARA's ability to capture repository context in code comprehension tasks (**RQ4**).

*1) Methodology:* To evaluate the code comprehension feature, we replicated a repository from the CodeSearchNet [20] dataset so that CLARA can consider repository context while comprehending a code file. Then, we selected 2 code files within this project, both included in our sampled set of 30 code comprehension examples for the users to analyze CLARA's code comprehension features (code file explanation summary and explanation of highlighted portion in code). For evaluating the code refactoring feature, we selected 2 code files from our 30 sampled code refactoring data from *MaRV* [21] dataset. For the code quality attribute detection feature, we also chose 2 code files from our sampled set of code files from the *Big-Vul* [22] dataset. To ensure a balanced evaluation, for each feature, we selected 2 code files (one where the model made good and accurate predictions, and one where the model underperformed) from our set of qualitatively evaluated samples.

In the study, participants were provided with detailed guidelines on the tool and its usage, along with a survey questionnaire. The survey questions were a combination of Likert-scale and open-ended questions and we provided participants with sample answers for reference.

*2) RQ1/RQ2 Results:* The participating users evaluated how useful and easy-to-use CLARA is.

<u>CLARA's Overall Usability:</u> Of 10 users, 8 found CLARA very easy to use, while 2 found the tool moderately easy to use (on a 5-point Likert scale, these options are the most positive). Additionally, 7 out of 10 users found CLARA's suggestions for 3 features "very clear and understandable" and 2 users found them "moderately clear and understandable" (2 most positive options on the Likert scale question, while 1 user was "neutral". Users also provided useful suggestions on improving CLARA's GUI.

<u>CLARA's Responsiveness:</u> On average, CLARA takes 10–12 seconds to respond to feature-related requests and chatbot queries. 7 participants found CLARA "very responsive" and 3 of them found the tool "moderately responsive" (the most positive options in the 5-point Likert scale)'.

<u>CLARA's Usefulness:</u> 7 out of 10 users found CLARA's features "very helpful" in exploring and analyzing open-source codebases, and 3 found CLARA "moderately helpful" in this regard (the most positive options in the Likert scale question). Additionally, 8 of the users "completely agreed" that the combination of CLARA's 3 features, along with the AI-assisted chatbot, simplifies the code exploration and analysis workflow, and 2 users "moderately agreed" in this regard (2 most positive Likert-scale options).

New Feature Suggestions: The participants suggested additional features for CLARA, including AI-assisted bug detection in code files and potential fixes, code documentation generation, and cross-file dependency & reference mapping.

*3) RQ3 Results:* : The users evaluated the quality and predictive accuracy of the suggestions of CLARA's 3 features.

Code Comprehension: All 10 participants agreed that CLARA provides clear, concise, and accurate suggestions when explaining entire code files as well as highlighted code fragments. Additionally, many of the users commented that CLARA's suggestions were very insightful and well-crafted.

Code Refactoring: 8 out of 10 participants agreed that CLARA provided useful and reusable refactored versions of the code, while one was unsure and one disagreed. The unsure and disagreeing users mentioned that CLARA's descriptive comments on the refactored segments were obscure, and in some cases, the refactored code did not handle exceptions properly. Overall, users valued this feature's potential usefulness.

Code Quality Attribute Detection: 7 out of 10 participants agreed that CLARA correctly detected the code quality attributes, while 3 were unsure. The reason behind their uncertainty was that LLM's predicted output varied slightly on different runs, which made the predictions slightly different even after giving the same input.

*4) RQ4 Results:* 6 out of 10 participants "completely agreed" and 4 of them "somewhat agreed" that CLARA properly captured the repository context in the code file explanation summary and accurately explained the highlighted code segment within the context of its code file.

## V. RELATED WORK

Few existing tools support individual code comprehension, analysis, and quality assessment tasks. For example, AskTheCode [8] is a custom GPT model that explains code snippets after providing repository URLs and custom prompts. BitoAI [9] and IntelliCodeEx [10] are IDE-based code comprehension tools that leverage LLMs to explain code after project configuration. AI-Code-Doctor [11] is a web app that analyzes a code file after copied and pasted into its interface. Additionally, some tools perform static analysis to detect code smells and vulnerabilities. For instance, JSNose [12] is a CLI tool that helps to detect code smells in JavaScript code, and CodeQL [13] is a query-based static analysis tool that detects vulnerabilities. Researchers have also proposed automated techniques for code smell [4, 5] and vulnerability detection [6, 7].

Existing code comprehension and analysis tools, as well as third-party generative AI models typically operate in isolation and require full repository downloads or manual prompt engineering to do code comprehension and analysis. Additionally, the existing tools have limited contextual understanding of the repository codebase and lack multi-feature support. In contrast, CLARA stands out as an easy-to-install, multi-featured browser extension that seamlessly integrates with GitHub repositories to assist users in code comprehension and analysis tasks without any prior configuration or manual prompting and generates context-aware responses.

## VI. CONCLUSIONS & FUTURE WORK

CLARA is a browser extension that assists users in code comprehension, refactoring, and code quality evaluation while navigating GitHub codebases. CLARA aims to minimize developers' effort and reduce their reliance on manual configurations. Evaluation results suggest that CLARA provides useful and context-aware suggestions. For future work, we plan to integrate a Retrieval-Augmented Generation (RAG) pipeline to CLARA provide more accurate and repository-oriented suggestions.